\documentclass{emulateapj}

\newcommand{\spi}{{\em Spitzer}}
\newcommand{\msx}{{\em MSX}}
\newcommand{\mum}{{\micron}}

\submitted{Submitted to the Astrophysical Journal Letters 29 Jul. 2005;
revised 18 Aug., 2005; accepted 25 Aug., 2005}
\journalinfo{ }

\begin{document}

\title{R CrB Candidates in the Small Magellanic Cloud: Observations of 
Cold, Featureless Dust with the {\em SPITZER} Infrared Spectrograph}

\author{
Kathleen E. Kraemer\altaffilmark{1}, 
G. C. Sloan\altaffilmark{2}, 
P. R. Wood\altaffilmark{3}, 
Stephan D. Price\altaffilmark{1}, 
Michael P. Egan\altaffilmark{1}
}

\altaffiltext{1}{Air Force Research Laboratory, Space Vehicles 
 Directorate, 29 Randolph Rd., Hanscom AFB, MA 01731; 
 kathleen.kraemer@hanscom.af.mil, steve.price@hanscom.af.mil, 
 michael.egan@osd.mil}

\altaffiltext{2}{Department of Astronomy, Cornell University, 
 108 Space Sciences Building, Ithaca, NY 14853; sloan@astro.cornell.edu}

\altaffiltext{3}{Research School of Astronomy \& Astrophysics, Mount 
 Stromlo Observatory, Weston Creek, ACT 2611, Australia; wood@mso.anu.edu.au}

\begin{abstract}
We observed 36 evolved stars in the Small Magellanic Cloud (SMC) using 
the low-resolution mode of the Infrared Spectrograph (IRS) on the 
{\em Spitzer Space Telescope}.  Two of these stars, MSX SMC 014 and 155, 
have nearly featureless spectral energy distributions over the IRS wavelength 
range (5.2--35 \micron) and F$_{\nu}$ peaking at $\sim$8--9 \micron. The data 
can be fit by sets of amorphous carbon shells or by single 600--700 K 
blackbodies. The most similar spectra
 found in extant spectral databases are of R CrB, although the 
spectral structure seen in R CrB and similar 
stars is much weaker or absent in the SMC sources. Both SMC stars 
show variability in the near-infrared. Ground-based 
visual spectra confirm that MSX SMC 155 is carbon-rich, as expected for 
R CrB (RCB) stars, and coincides with an object previously identified as an 
RCB  candidate. The temperature of the underlying star is lower for MSX SMC 
155 than for typical RCB stars. The strength of the C$_2$ Swan bands and the 
low temperature suggest that it may be a rare DY Per-type star, only the 
fifth such identified. MSX SMC 014 represents a new RCB candidate
in the SMC, bringing the number of RCB candidates in the SMC to six. It is
the first RCB candidate discovered with \spi\ and the first identified 
by its infrared spectral characteristics rather than its visual variability.

\end{abstract}


\section{Introduction \label{sec.intro}} 

R Coronae Borealis (RCB) stars are a rare type of hydrogen-deficient, 
carbon-rich, evolved star that undergo
strong, irregular brightness variations in the optical \citep[see][for a 
detailed review]{cla96}. Only about 50 or so
RCB stars or RCB candidates have been identified in the Milky Way, less than 
a score in the Large 
Magellanic Cloud (LMC) \citep{alc96,alc01}, and five in the Small 
Magellanic Cloud (SMC) \citep{mhcc03,tis04}. Distances to  Galactic 
RCB stars are problematic, and thus distant-dependent properties of these 
poorly-understood objects, such as absolute luminosity, are also problematic.
Distances to the LMC and SMC, in contrast, are well constrained. 
Identification and observation of RCB candidates in the LMC and SMC are 
therefore 
essential for characterizing and understanding this phase of stellar evolution.

\section{Observations \label{sec.obs}} 

\begin{figure}
\includegraphics[width=3.5in]{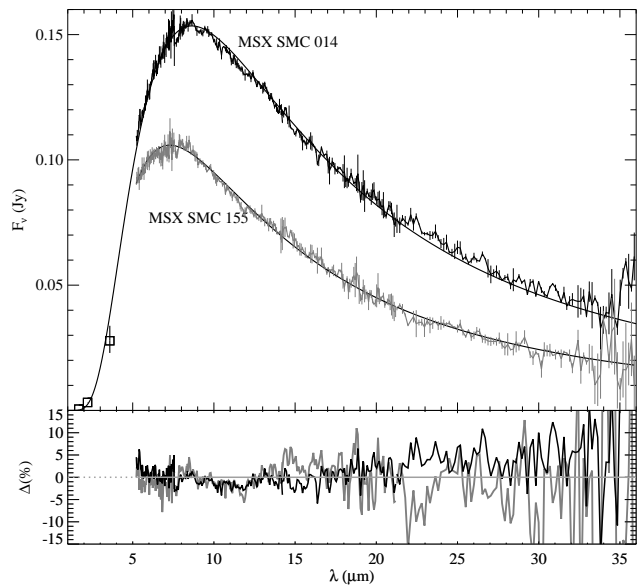}
\caption{Top: IRS full-range spectra for MSX SMC 014 (black) and 155 (gray), 
including 
uncertainties. Smooth curves show
the blackbody functions fit to the IRS data: 590 K for MSX SMC 014 and  700 K
for MSX SMC 155 (\S 4). Bottom:
 residuals (\%) between the data and the blackbody fits. The gray
line shows the residuals for MSX SMC 155 and the black line for MSX SMC 014.
The squares are the near-IR SSO data for MSX SMC 014 (see text).}
\label{fig.bbs}
\end{figure}

We observed MSX SMC 014 and 155\footnote{Following the nomenclature of 
\citet{evp01}. In the \msx\ Point Source Catalog V2.3 
\citep{epk03},
 MSX SMC 014 = G303.4131$-$42.9358 and MSX SMC 155 = 
G302.3214$-$44.4109.}, two evolved stars in the SMC, with the Infrared 
Spectrograph \citep[IRS;][]{hou04} on the {\it Spitzer Space Telescope} 
\citep{wer04} on 2004 October 25. The 2MASS positions (J2000) 
are (00 46 16.33, $-$74 11 13.6) and (00 57 18.15, $-$72 42 35.2) for MSX SMC 
014 and 155, respectively. The observations used  the Short-Low (SL) 
and Long-Low 
(LL) modules, which have a wavelength range of 5.2--35~\mum\ and spectral
resolution of $\sim$100.  We extracted the spectra
from the flatfielded images produced by the \spi\ Science
Center pipeline S11.0 using the software available with the \spi\
IRS Custom Extractor. The spectra from each order 
and nod were extracted separately
and flux calibrated with HR 6348 
for SL, and HR 6348, HD 166780, and HD 173511 for 
LL (see Sloan et al. in preparation). Errors were estimated by
comparing the nods, and multiplicative corrections were used to
remove discontinuities between orders caused by random pointing
offsets.  Figure \ref{fig.bbs}  shows the resulting, featureless 
spectra, which look remarkably like blackbody curves.

\begin{deluxetable*}{ccccc}
\tablecaption{Near-IR Photometry\label{tab.best}}
\tablewidth{0pt}
\tablehead{ \colhead{2004 Nov 25} & \colhead{J} & \colhead{H} & 
  \colhead{K} & \colhead{L} }
\startdata
\\
MSX SMC 014 & \nodata          & 15.683$\pm$0.093 mag & 
              13.212$\pm$0.021 & 9.939$\pm$0.211 \\
MSX SMC 155 & 13.740$\pm$0.064 & 12.548$\pm$0.010     & 
              11.370$\pm$0.016 & 9.144$\pm$0.090 \\[0.8mm]
\hline
\hline
\\
K Band      & 2004 Nov 25      & 2005 Jan 26          & 
              2005 Mar 14      & 2005 Jul 24 \\[0.8mm]
\hline
\\
MSX SMC 014 & 13.212$\pm$0.021 & 13.362$\pm$0.021     & 
              13.284$\pm$0.052 & 13.493$\pm$0.024 \\
MSX SMC 155 & 11.370$\pm$0.016 & 11.522$\pm$0.014     & 
              12.003$\pm$0.024 & 12.327$\pm$0.012
\enddata
\end{deluxetable*}

Optical spectra of MSX SMC 155 and two comparison stars were 
taken on 2004 November 22 (28 days after the IRS data)
using the Double Beam Spectrograph on the 2.3 m
telescope of the Australian National University at Siding Spring 
Observatory (SSO).
The spectral range was 0.45--1.05 $\mu$m and the resolution was
10 \AA.  The data were reduced with standard IRAF procedures using
HR 718 as a flux standard and the weak-lined giant HD 26169 to remove
telluric features. Due to some cloudiness during the observations, 
the relative flux calibration is correct but the absolute flux is uncertain.
In addition to the optical spectra, JHKL photometry was taken of both stars on
2004 November 25 using the IR imager CASPIR on the 2.3 m telescope. They 
have since been monitored at K band 
at roughly two--four month intervals (Table 1).

\section{R CrB Star Candidates} 

Evolved stars can generally be divided into oxygen-rich and carbon-rich groups,
with oxygen-rich stars usually showing mid-IR silicate or alumina
dust features, and carbon stars showing features from SiC dust and PAHS, and 
bands from C$_2$H$_2$ and other carbon-bearing molecules.  In contrast, the 
mid-IR spectra of MSX SMC 014 and 155 are virtually featureless, offering 
few clues as to the chemistry of their dust, let alone their photospheres.

\begin{figure}
\includegraphics[width=3.5in]{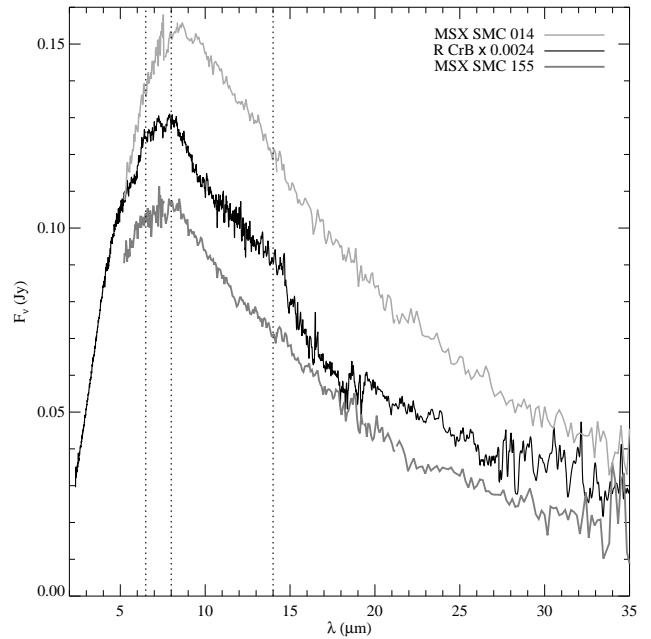}
\caption{The top trace (light gray) shows  MSX SMC 014; 
the middle trace (black) shows  R CrB \citep{skps03}, scaled
by 0.0024; the bottom trace (dark gray) shows  MSX SMC 155.
The dotted lines indicate the regions of possible spectral features: 
amorphous carbon at 6 and 14 \micron\ in R CrB \citep{lrpi01} and a possible, 
unidentified feature near 8 \micron\ in all three.} 
\label{fig.rcbcomp}
\end{figure}

We examined spectra from the Short Wavelength Spectrometer (SWS) and 
the Photo-Polarimeter on the {\em Infrared Space Observatory (ISO)}, 
looking for similarly bland spectral energy distributions (SEDs) where 
F$_{\nu}$ peaks $\sim$7--9 \micron, which 
led us to spectra in the 3.W  class of \citet{kspw02}. Closer inspection 
revealed that the objects in 3.W 
without silicate absorption are mostly RCB stars (those with 
silicate absorption are usually Wolf-Rayet stars). Figure \ref{fig.rcbcomp} 
compares the  SWS spectrum of 
R CrB itself and the IRS spectra of MSX SMC 014 and 155. R CrB shows 
broad emission features around 6 and 14 \micron\ identified by 
\citet{lrpi01} as amorphous carbon features. 
The spectra of the SMC stars agree quite well with the R CrB data, although
they have at best very weak features at 6 and 14 \micron. The lack of 
hydrogen in RCB stars naturally leads to the formation of amorphous
carbon dust \citep[e.g.][]{hhd84}, which is nearly featureless in the IR, 
instead of hydrogen-rich particles such at PAHs, which have strong IR 
features that are not present in the SMC data. 

\begin{figure}
\includegraphics[width=3.5in]{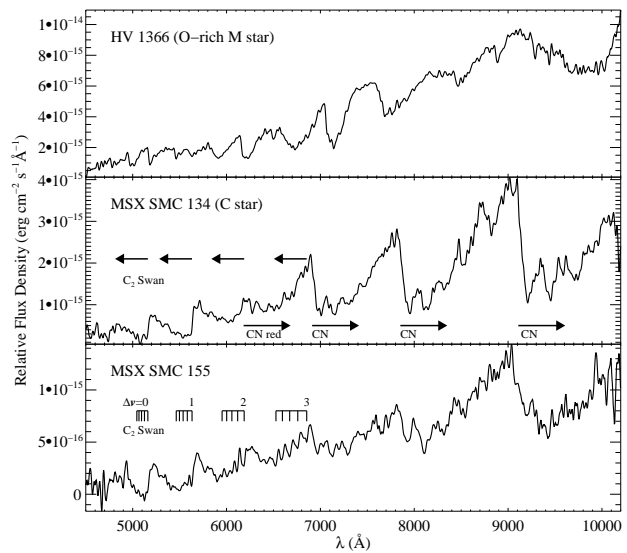}
\caption{The visual spectrum of MSX SMC 155 (bottom) is a much better match 
to the carbon star MSX SMC 134 (middle) than the O-rich M star HV 1366 (top).
The C$_2$ Swan bands and CN red bands are indicated on the MSX
SMC 134 plot. Individual vibrational bandheads for the Swan bands are shown
in the MSX SMC 155 panel, which has particularly strong $\Delta v=3$ bands.  
}\label{fig.vis}
\end{figure}

\begin{figure}
\includegraphics[width=3.5in]{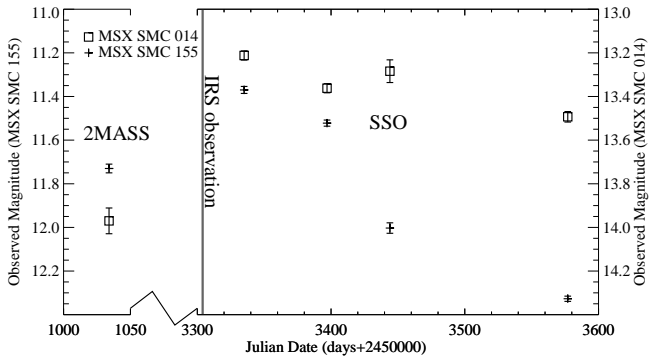}
\caption{K-band observations of MSX SMC 155 (pluses) and MSX SMC 014 (boxes).
The first set of points are the 2MASS K$_s$ data; the rest are the 
more recent observations from the 
SSO (note the break in x-axis
scale). The y-scale on the left side corresponds to 
MSX SMC 155 and that on the right side is for MSX SMC 014. Error bars are 
1$\sigma$ measurement uncertainties. The thick gray line indicates the date of
the IRS observations (JD 2453304).
}\label{fig.var}
\end{figure}

Most (if not all) RCB candidates have previously been identified by their
strong and irregular variability in the visible. \citet{tis04}
noted such variability in MSX SMC 155 and identified it as a 
candidate RCB star based on the 5--6 mag variation in
 EROS 2 data (Exp\'erience de Recherche d'Objets 
Sombres)\footnote{EROS name: J005718-724235(sm0067m2813b); MACHO identifier:
207.16426.1662}. 
We confirmed the carbon-rich nature of the central star of MSX SMC 155 with 
visual spectroscopic observations, as Figure \ref{fig.vis} shows. 
Our multi-epoch K-band data, shown in Figure \ref{fig.var} along with the 
2MASS K$_s$ datum from 1998, indicate that MSX SMC 155 is variable in the
near-IR as well as the visible. The timescale and magnitude of the K-band 
variation in MSX SMC 155 (one magnitude decline in 240 days) is fairly
typical of RCB stars \citep{fcr97} as are the near-IR colors \citep{f97}.
Thus, our visual and IR observations strongly support the candidacy of 
MSX SMC 155 as an RCB star. 

MSX SMC 014 does not appear in any of the databases produced by EROS or 
other dark matter searches that target the Magellanic Clouds 
 \citep[e.g.][]{alc96a,sz05}, being
outside their observed fields. Due to its dimness, we were 
unable to confirm its carbon-rich nature with 
ground-based spectroscopy. We did detect K-band variability (Fig. 
\ref{fig.var}), as with MSX SMC 155. The recent variation is not as great as
MSX SMC 155 showed (0.96 mag vs. 0.28 mag) but there has been a $\sim$0.7 mag
increase in brightness since the 2MASS observation in 1998. 
It appears that both MSX SMC 155 and 014 were near maximum K brightness
when observed by \spi, with MSX SMC 155 starting a fading event about 30--60
days after the \spi\ observation (Fig. \ref{fig.var}). MSX SMC 014 was redder
than the stars studied by \cite{fcr97} but it is similar to the LMC RCB star
HV 2379 during its faded period \citep{bw83}.
Given the similarity of the mid-IR spectrum of MSX SMC 014 to those of 
MSX SMC 155 and R CrB, as well as the near-IR variability, we feel confident 
in identifying it as the sixth RCB candidate in the SMC. It is also the first 
identified by its IR spectrum (as opposed to visual variability) 
and the first identified by \spi.

\section{Dust and Stellar Properties\label{sec.dust}} 

The near-featurelessness of the spectra precludes a significant 
contribution from grains with strong spectral structure, such as graphite or 
silicates which dominate the mid-IR dust spectra of ordinary  evolved stars. Thus, we initially fit single temperature Planck 
functions to the spectra, which are shown in Figure \ref{fig.bbs}. 
MSX SMC 014 can be fit with curves in the temperature range 580--610 K,
with the best fit at 590$\pm$5 K. MSX SMC 155 is best fit with a 
700$\pm$5 K blackbody but could be as cool as 690 K. The residuals left after 
subtraction of the indicated blackbody from the IRS data are 
less than 10\% for most of the IRS wavelength range (Fig. \ref{fig.bbs}). 
Dust temperatures for Galactic RCB stars typically lie in the range 600--900 
K \citep{kw84,hjw85},
so the single-component dust temperatures of the SMC objects are comparable 
to the Galactic population.

\begin{figure}
\includegraphics[width=3.5in]{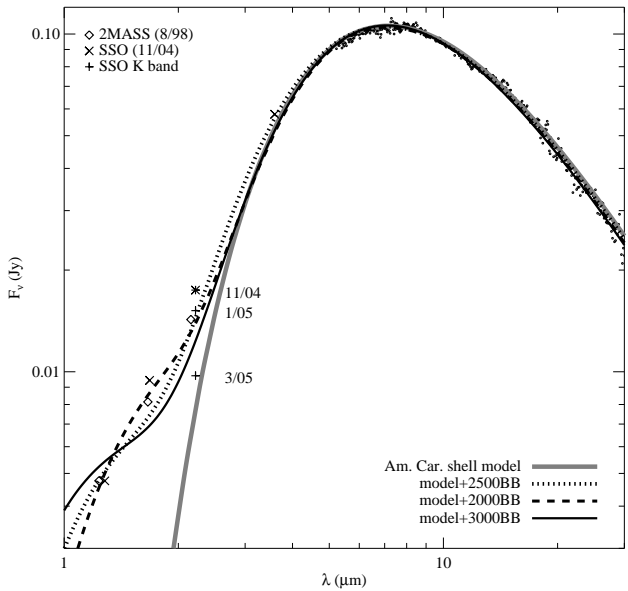}
\caption{IRS spectrum (small circles) for MSX SMC 155 plus the 2MASS data 
(diamonds), the SSO (crosses) data, and the multi-epoch K-band data from the 
SSO (pluses with dates). 
The thick gray line shows the model for a 2500 K central star with 12 
shells of amorphous carbon starting at R=6.1R$_*$. The dotted line adds
 a small contribution from a 2500 K blackbody. The dashed 
line shows a similar shell+blackbody model but for a 2000 K star, and the black
line is for a 3000 K star.}
\label{fig.bblog}
\end{figure}

Amorphous carbon,  relatively featureless in the mid-IR as noted above, 
is a likely carrier, so we obtained optical constants from 
\cite{suh00} and \cite{zub96}\footnote{The optical constants are 
available from their respective websites 
http://ast.chungbuk.ac.kr/$\sim$kwsuh/d-opt.htm and 
http://idlastro.gsfc.nasa.gov/$\sim$zubko/nk.html}.
Although mostly featureless, the amorphous carbon SEDs are significantly 
narrower than the observed spectra, if
the grains are assumed to be at a single temperature (i.e., at a single
distance from the central star). A simple model, though, with a set of
SEDs from amorphous carbon ``shells'' at varying 
distances from the central star can fit
the IRS data as well as the single-temperature blackbodies. (Note that the
shells are simple representations of the irregular dust condensations that
likely cause the visible obscuration and are not intended as actual physical 
entities.) 
The temperature range for the dust shells is about 500--1100 K (Fig. 
\ref{fig.bblog}). However, the 
IRS data do little to constrain the stellar temperature 
since, for instance, a hotter star can be compensated for by placing the 
first shell at a greater distance from the star. 

The near-IR data, though, do provide
constraints on the stellar temperature and, to a lesser extent, the position
of the inner shell, its visual opacity, and the number of shells. 
(Complete modeling to explore the full parameter space is deferred 
to a future paper.) For MSX SMC 014, the K- and L-band points lie on 
the 590 K blackbody (Fig. 1) or multi-shell amorphous carbon curves. The  
H-band point may be slightly above the curves which may indicate tentative 
detection of a photospheric contribution (the J band is an upper limit in 
both the SSO and 2MASS data) but by itself is insufficient to constrain the 
models. The near-IR data for MSX SMC 155, though, lie well above the 
extrapolated mid-IR curves, so we do detect a photospheric contribution. 
For the amorphous carbon shells, a small contribution from 
stellar temperatures of $\sim$2000--2500 K is supported by the near-IR data;
higher temperatures produce too much J-band flux but underestimate the 
K$_s$ band flux. For the single-temperature dust model, slightly higher stellar
temperatures fit although this is a less physically plausible model.
Figure \ref{fig.bblog} shows typical results for the fitted models.
The ranges in stellar temperature estimated 
for MSX SMC 155 are colder than even the unusually cool RCB star DY Per 
\citep[$\sim$3500 K,][]{kb97}.

Comparison of the optical spectrum of MSX SMC 155, particularly the 5635 \AA\
band of C$_2$, with the catalog of C star spectra of \cite{bsk96}
 and the $T_{eff}$ values given for the catalog stars by \cite{bkr01}
suggests that $T_{eff}$ for MSX SMC 155 is most likely $\sim$3000--4000 K, 
although cooler values down to 2500 K cannot be ruled
out. The strong $\Delta v=3$ vibrational bandheads of C$_2$ in MSX SMC 
155 ($\lambda\lambda$ 6526--6854 \AA)  suggest that C$_2$ has been excited
to unusually high vibrational levels. These bandheads are weak or absent in
typical C stars \citep[e.g. MSX SMC 134 in Fig. \ref{fig.vis};][]{bsk96,alc01} 
but were also strong in HV 2379 \citep{bw83}. Finally, our visual spectrum is
quite similar to those of the four LMC RCB stars of \cite{alc01} 
which have 4000--7000 \AA\ spectra resembling DY Per. We suggest that MSX SMC 
155, too, may be a cool, DY Per-type RCB star. 

The distance to the SMC is known (60 kpc or a distance modulus of 
18.9), so we can  calculate the luminosities. We integrated under the 
single-temperature (590 K) blackbody
fit for MSX SMC 014 and a model of amorphous carbon shells plus a 2500 K
blackbody (dotted line in Fig. \ref{fig.bblog}) for MSX SMC 155. 
These gave luminosities of L$\approx$9700 L$_\sun$ and L$\approx$11000 L$_\sun$
of which roughly 70\% and 40\% emerge within the IRS bandpass for  
MSX SMC 014 and 155, respectively. Given these luminosities and the 
cool photospheric temperatures 
suggested for the central stars, it appears that these objects may still be
on the asymptotic giant branch (AGB) but are undergoing the same type of 
episodic
carbon shell ejections that are seen in warmer RCB stars.

\section{Summary \label{sec.conc}} 

We observed two objects in the SMC with largely featureless spectra 
in the mid-IR.
These spectra are most similar to the SWS data for R CrB and other Galactic 
RCB stars. MSX SMC 155 has been identified based on its 
optical variability as one of five RCB candidates in the SMC.  The second 
star, MSX SMC 014, has no readily available visual photometry,
but our K-band observations, along with the 2MASS data, indicate it is 
a near-IR variable. Based on the similarity of its IRS spectrum to 
that of MSX SMC 155, we identify MSX SMC 014 as the first RCB candidate
found by its IR spectral properties rather than by optical variability, and
also as the first RCB candidate discovered by \spi.  The spectra can be fit by
either a single-temperature blackbody or a set of amorphous carbon shells 
produced by different episodes of dust formation. Calculated luminosities and
estimated stellar temperatures imply that the central star for
MSX SMC 155 may be a carbon star still on the AGB. We suggest that it
 is the first DY Per star, a rare type of very cool RCB star, 
identified in the SMC.

\acknowledgements
We wish to thank Bill Forrest for useful discussions on amorphous
carbon shells and  the referee, Geoffrey Clayton, whose helpful
suggestions improved the paper. This work is based in part on observations 
made with the {\em Spitzer Space Telescope}, which is operated by JPL/Caltech
 under NASA contract 1407. Support for this work was 
provided in part by NASA; PRW received funding for this work from the 
Australian Research Council. This research has made use of NASA's 
Astrophysics Data System, 2MASS, and Simbad.

\end{document}